\documentclass[runningheads]{llncs}
\usepackage{graphicx}
\usepackage[linesnumbered,ruled]{algorithm2e}
\usepackage{tabularx}
\usepackage{amsmath}
\usepackage{xcolor}
\usepackage{makecell}
\usepackage[hyphens]{url}
\usepackage[hidelinks]{hyperref}
\hypersetup{breaklinks=true}
\usepackage{cite}
\usepackage{tabularx}
\begin{document}
\title{TensorFlow as a DSL for stencil-based computation on the Cerebras Wafer Scale Engine}

\titlerunning{TensorFlow as a DSL on Cerebras WSE for stencil-based computation codes}
%


\author{Nick Brown\inst{1} \and
Brandon Echols\inst{2} \and
Justs Zarins\inst{1} \and
Tobias Grosser\inst{3}}
\authorrunning{N. Brown et al.}
%
\institute{EPCC, University of Edinburgh, Bayes Centre, Edinburgh, UK \and
Lawrence Livermore National Laboratory, Livermore, California, USA \and
School of Informatics, University of Edinburgh, Informatics Forum, Edinburgh, UK}
\maketitle              
\begin{abstract}
The Cerebras Wafer Scale Engine (WSE) is an accelerator that combines hundreds of thousands of AI-cores onto a single chip. Whilst this technology has been designed for machine learning workloads, the significant amount of available raw compute means that it is also a very interesting potential target for accelerating traditional HPC computational codes. Many of these algorithms are stencil-based, where update operations involve contributions from neighbouring elements, and in this paper we explore the suitability of this technology for such codes from the perspective of an early adopter of the technology, compared to CPUs and GPUs. Using TensorFlow as the interface, we explore the performance and demonstrate that, whilst there is still work to be done around exposing the programming interface to users, performance of the WSE is impressive as it out performs four V100 GPUs by two and a half times and two Intel Xeon Platinum CPUs by around 114 times in our experiments. There is significant potential therefore for this technology to play an important role in accelerating HPC codes on future exascale supercomputers.
\end{abstract}
\section{Introduction}

Scientists and engineers are forever demanding the ability to model larger systems at reduced time to solution. This ambition is driving the HPC community towards exascale, and given the popularity of accelerators in current generation supercomputers it is safe to assume that they will form a major component of future exascale machines. Whilst GPUs have become dominant in HPC, an important question is the role that other more novel technologies might also play in increasing the capabilities of scientific simulation software. One such technology is Cerebras' Wafer Scale Engine (WSE) which is an accelerator containing hundreds of thousands of relatively simple, AI, cores. Whilst the major target for Cerebras to this point has been accelerating machine learning workloads, as the cores are optimised for processing sparse tensor operations this means they are capable of executing general purpose workloads, and furthermore combined with massive on-chip memory bandwidth and interconnect performance. Put simply, the WSE has significant potential for accelerating traditional HPC computational kernels in addition to machine learning models.

There are currently a handful of Cerebras machines which are publicly available, making testing and exploration of the architecture difficult. Furthermore, the software stack is optimised for machine learning workloads, and whilst Cerebras are making impressive progress in this regard, for instance the recent announcement of their SDK \cite{sdk}, at the time of writing machine interaction is commonly driven via high level machine learning tools. It is currently a very exciting time for the WSE, with Cerebras making numerous advances in both their software and future hardware offering. Consequently, whilst the technology is still in a relatively early state, at this stage understanding its overall suitability for HPC workloads compared with other hardware is worthwhile, especially as the Cerebras offering is set to mature and grow in coming years.

In this paper we explore the suitability of the Cerebras WSE for accelerating stencil-based computational algorithms. Section \ref{sec:bg} introduces the background to this work by describing the WSE in more detail and how one interacts with the machine, along with other related work on the WSE. In Section \ref{sec:programming} we explore how one must currently program the architecture for computational workloads and then, by running on a Cerebras CS-1, in Section \ref{sec:results} use a stencil-based benchmark to compare the performance properties of the WSE against four V100 GPUs and two 18-core Intel Xeon Platinum CPUs, before concluding in Section \ref{sec:conclusions}.

\section{Background and related work}
The Cerebras WSE has been used by various organisations, including large global corporations, for accelerating machine learning. Already there have been numerous notable successes from running AI models on the WSE including new drug discovery \cite{drug-cs1}, advancing treatments for fighting cancer \cite{pendse2020memory}, and helping to tackle the COVID-19 pandemic \cite{trifan2021intelligent}. The benefits of accelerating machine learning workloads has been well proven, however there are far fewer studies concerned with using the WSE to run more traditional computational tasks.

One such study was undertaken in \cite{rocki2020fast} where the authors ported the BiCGSTAB solver, a Krylov Subspace method for solving systems of linear equations, and also a simple CFD benchmark onto the Cerebras CS-1. Whilst their raw results were impressive, the authors used Cerebras' low level interface for this work, programming each individual core separately and manually configuring the on-chip network. This required a very deep understanding of the architecture, and furthermore as the work was undertaken in part by Cerebras employees they had access to this proprietary tooling which is not publicly available to users. 

In this work we focus on stencil-based algorithms because of their suitability for mapping to the WSE architecture and TensorFlow programming interface (see Section \ref{sec:programming}). When calculating the value of a grid cell stencils represent a fixed pattern of contributions from neighbouring elements. Most commonly operating in iterations, at each iteration the value held in each grid cell will be updated based upon some weighted contribution of values held in neighbouring cells. This form of algorithm is widespread in scientific computing and hence represents the underlying computational pattern in use by a large number of HPC codes.

\label{sec:bg}
\subsection{Cerebras Wafer Scale Engine}
The Cerebras Wafer Scale Engine (WSE) is a MIMD accelerator and on the CS-1, the hardware used for this work, there are approximately 350000 processing cores running concurrently and able to executing different instructions on different data elements. The WSE provides more flexibility than a GPU, for instance, where on that accelerator groups of cores must operate in lock-step within a warp. At the physical level the WSE is composed of a wafer containing 84 dies, with each die comprising 4539 individual tiles. Each tile holds a single processing element, which is a computational core, a router, and 48KB of SRAM memory. In total there is approximately 18GB of SRAM memory on the CS-1 but this is distributed on a processing element by processing element basis. Each computational core supports operations on 16-bit integers, and both 16-bit and 32-bit floating point numbers, with the IEEE floating point standard supported for both floating point bit sizes and additionally Cerebras's own CB16. Each core provides 4-way SIMD for 16-bit floating point addition, multiplication, and fused multiply accumulate (FMAC) operations, 2-way SIMD for mixed precision (16-bit multiplications and 32-bit additions), and one operation per cycle is possible for 32-bit arithmetic. 


The WSE is designed to accelerate computation involved in model training and inference, with numerous support functions undertaken by the host machine. The host is connected to the WSE via twelve 100 GbE network connections, and undertakes activities include model compilation, input data preprocessing, streaming input and output model data, and managing the overall model training. The Cerebras machine used for this work is a CS-1 hosted by EPCC and connected to a host Superdome Flex Server (containing twenty four Intel Xeon Platinum 8260 CPUs, with each CPU containing 24 physical cores and a total of 17TB RAM).

\subsection{Programming the Wafer Scale Engine}
\label{sec:programmingwse}
In \cite{rocki2020fast} the authors programmed their kernels for the CS-1 using a bespoke low level interface, however this is proprietary and not exposed to users. Cerebras have recently announced the availability of their SDK \cite{sdk} for general purpose programming of the WSE and whilst this is a very important step in widening the workloads that can be executed on the architecture, it requires an investment of time for programmers to gain the expertise in order to be able to write optimal code for the WSE using it. Consequently in this work we use the TensorFlow API, which abstracts the tricky and low level details of decomposing the workload into tasks, mapping these to cores, and determining the appropriate routing strategy. Hence whilst our objective is to focus on stencil-based, rather than machine learning, codes, by encoding our algorithm via TensorFlow it enables us to undertake performance explorations for this workload, to understand whether it is worthwhile investing the time in using the Cerebras SDK, and also means that such algorithms can be ported to the WSE more quickly to undertake such evaluations.

The WSE supports a subset of TensorFlow functionality, and in this work we use two major building blocks to encode stencil-based algorithms. The first building block are dense layers, which are fully-connected meaning that every value provided as an input to the layer will have a connection to every output value of the layer. As such the operation performed by a dense layer is a matrix-matrix multiplication with a batch of input tensors and weight matrix resulting in, for every output value, each input value multiplied by a specific weight and intermediate values added together to form the result. 


The second TensorFlow construct used in this work are convolution layers, where a kernel slides across the input tensor and performs a convolution product to calculate results. For each element of the output, the kernel weight values will be multiplied with a subset of the input values. In the 2D case, the filter can be thought of as sliding from left to right and up to down, and whilst TensorFlow includes convolution layers that operate in one, two, and three dimensions, at the time of writing the Cerebras software stack only supports the 2D convolution layer. In this \emph{Conv2D} layer the data-structure is comprised of four dimensions which are the batch size, number of channels (the depth of the input tensor, for instance red, green, blue for an image), rows, and columns. Whilst the WSE provides single and half precision in hardware, the Cerebras software stack only supports mixed precision (single and half) at the TensorFlow API level.

\section{TensorFlow for encoding stencil-based algorithms on the Wafer Scale Engine}
\label{sec:programming}
In this work our objective has been to implement a stencil-based benchmark and for this we selected the Jacobi iterative method for solving Laplace's equation for diffusion in multiple dimensions. Whilst this is a fairly simplistic solver compared to the BiCGSTAB method explored on the CS-1 in \cite{rocki2020fast}, the limitation of having to encode the algorithm via TensorFlow imposes some limitations. Furthermore, the underlying computational pattern is similar and represents an important class of algorithms and solvers. Consequently insights obtained from this benchmark on the WSE are highly relevant and interesting to the wider HPC community. Other benchmarks, such as the Open Earth Compiler benchmark suite \cite{gysi2021domain}, were considered however they were not readily representable in TensorFlow in a form that would build with the Cerebras software stack. 

The first approach we explored used a dense layer to undertake the Jacobi stencil computation. A sketch of this algorithm is illustrated in Algorithm \ref{alg:stencil_dense_calc}, where \emph{x} is the input tensor containing data being operated upon, and \emph{stencil} is a matrix representing the stencil operation. The input tensor is first flattened and then, along with \emph{stencil}, passed to the \emph{Dense} TensorFlow layer which will undertake the calculation. This operation is repeated \emph{iterations} times.

\begin{algorithm}[ht]
\caption{Stencil Calculation with Dense Layer}
\label{alg:stencil_dense_calc}
    \SetKwInOut{Input}{Input}
    \SetKwInOut{Output}{Output}
    
    \underline{function model-function} $(x, stencil, iterations, N)$\;
    \Input{x - the input tensor for the stencil calculations\\
    stencil - matrix used by Dense layer to perform stencil calculation\\
    iterations - the number of times the calculation will be performed\\
    N - total number of elements per step}
    \Output{result of performing stencil calculation on input tensor}
    $values = Flatten(x)$\;
    \For{$i\gets0$ \KwTo $iterations$}{
        $values = Dense(N, kernelInitializer=stencil)(values)$\;
    }
    
    \Return $values$
\end{algorithm}

$N$ is the total size of the input tensor per step, \emph{x}, which is of size equal to $X$ in one dimension, $X*Y$ in two dimensions, and $X*Y*Z$ in three dimensions. TensorFlow drives the dense layer with inputs over many steps, and the overarching problem size being operated upon is $N * number of steps$. The problem is therefore decomposed into tiles each of size $N$, and overlapping is undertaken to ensure boundary neighbours from one tile are available to another. This decomposition of the problem into steps, each of size $N$, is required to fit the hardware's memory and compute limits.

There are several advantages to programming the WSE using dense layers such as the ability to readily handle any number of input dimensions because the input is flattened regardless. Furthermore, because we explicitly define the stencil calculation then special cases, such as non-zero boundary conditions, can be handled without the need for conditional statements or other operations. For instance in this example the stencil matrix value can be set to 1 in order to maintain boundary conditions throughout the calculation. 

\begin{figure}[h]
\centering
\includegraphics[scale=0.65]{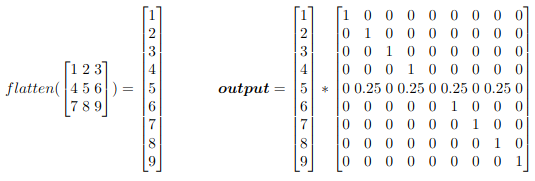}
\caption{Illustration of dense layer operations for solving Laplace's equation for diffusion in 2D with $X=Y=3$}
\label{fig:dense_layer}
\end{figure}

However, the major disadvantage of this approach is that the dense layer is of size $N^2$ (where \emph{N} is the total size of the input tensor per step). Depending upon the equation being solved this can involve a significant amount of redundant storage and computation. Figure \ref{fig:dense_layer} provides an illustration for solving Laplace's equation for diffusion in 2D with $X=Y=3$. This is first flattened into a vector of size $N=X*Y=9$ and then a matrix-vector product undertaken to calculate the results. In this example all cells on the boundaries, which is every element apart the middle value, \emph{5}, remains unchanged which corresponds to \emph{1} in the stencil matrix as it is a boundary condition. The \emph{0.25} values in the stencil matrix average neighbouring values, with every other element a zero and not contributing to the result. However these zeros must still be stored in the matrix and computations undertaken with them on them regardless.

Another approach, as introduced in Section \ref{sec:programmingwse}, is to use a convolution layer where the stencil is represented as a much smaller data window that slides across the input values. A sketch of the code for driving the convolution layer approach is illustrated in Algorithm \ref{alg:stencil_conv_calc} where, in contrast to the dense layer of Algorithm \ref{alg:stencil_dense_calc}, input values are not flattened because the convolution layer is dimensioned. Furthermore, there are two additional arguments, \emph{dataFormat} and \emph{padding} provided to this layer at line 3. The former determines the ordering of the dimensions in the input and output tensors, and the CS-1 only supports \emph{channelsFirst}. The second option ensures that the output is the same shape as the input by undertaking additional padding if required, where \emph{same} results in padding with zeros evenly to the left/right or up/down of the input. 

\begin{algorithm}[ht]
\caption{Stencil Calculation with Convolution Layer}
\label{alg:stencil_conv_calc}
    \SetKwInOut{Input}{Input}
    \SetKwInOut{Output}{Output}
    
    \underline{function model-function} $(x, stencil, iterations, stencilShape)$\;
    \Input{x - the input tensor for the stencil calculations\\
    stencil -  filter for the Conv2D layer performing stencil calculation\\
    iterations - the number of times the calculation will be performed\\
    stencilShape - the shape of the stencil}
    \Output{result of performing stencil calculation on input tensor}
    
    \For{$i\gets0$ \KwTo $iterations$}{
        $x = Conv2D(1, stencil, kernelInitializer=stencilInit, dataFormat='channelsFirst', padding='same')(x)$\;
    }
    \Return $x$
\end{algorithm}

\begin{figure}[h]
\centering
\includegraphics[scale=0.90]{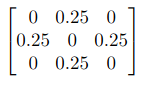}
\caption{Illustration of convolution layer kernel for 2D Laplace's equation for diffusion}
\label{fig:convolution_layer}
\end{figure}

The major benefit of the convolution layer is that, because the defined filter slides across the input, it decouples the size of the stencil matrix from the input tensor size. The convolution layer stencil for the same Laplace's equation for diffusion in 2D is illustrated in Figure \ref{fig:convolution_layer}, where irrespective of the input tensor size, $N$, nine values are required for the 2D case. Consequently, whilst there are some zeros still present, representing wasted storage and computation, their number is very significantly reduced in comparison to the dense layer approach. 

However there are two disadvantages with using convolution layers as sketched in Algorithm \ref{alg:stencil_conv_calc}, firstly stencil-based algorithms with non-zero boundary conditions are not possible because padding adds extra zero elements. To enable boundary condition values other than zero, the padding of the convolution layer must be changed to mode \emph{valid}, with the algorithm then manually defining the padding of the input. The most convenient approach to do this would be to use the \emph{tensorflow.pad} operation, which pads the outer edge with zeros, and boundary conditions could then be added around this padded input, driven by a concatenate layer. However, at the time of writing, both the pad operation and concatenate layer are not supported by the Cerebras software stack. 

Instead a mask must be created that will zero out the edges that were updated by the convolution layer and then subsequently add the boundary conditions back in. The mask is a tensor of the same shape and size,  $N$, as the input tensor and contains \emph{1} in the internal values and \emph{0} on the outer, boundary condition, locations. Multiplying the mask by the output zeros out the boundary conditions and then a further, \emph{boundary conditions} tensor which holds zeros for inner elements and the boundary conditions themselves, is added to the masked intermediate result. Whilst this approach is not ideal, as it results in additional runtime overhead, it is required because the Cerebras software stack does not yet fully support the entire TensorFlow API which would enable better alternatives.

\begin{figure}[h]
\centering
\includegraphics[scale=0.40]{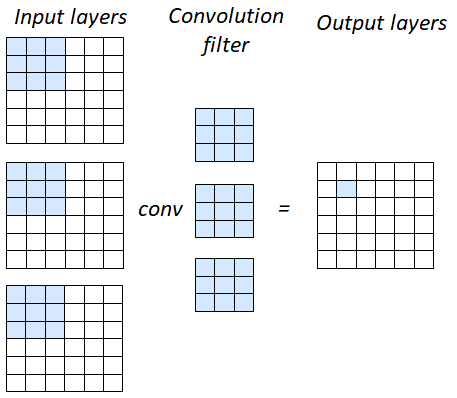}
\caption{Illustration of 3D convolution approach with the input in 3D but output in 2D}
\label{fig:convolution_layer_out_2d}
\end{figure}

\begin{figure}[h]
\centering
\includegraphics[scale=0.35]{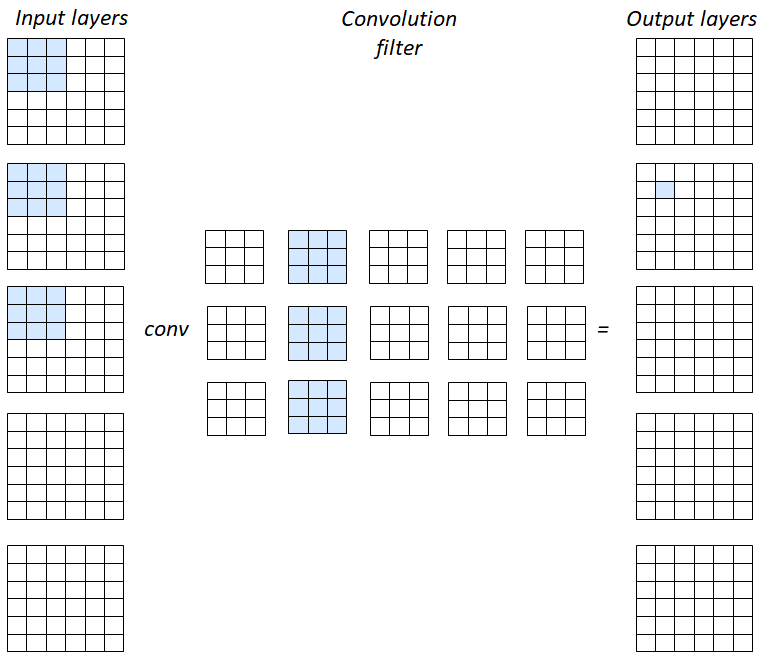}
\caption{Illustration of 3D convolution approach with the input and output in 3D}
\label{fig:convolution_layer_out_3d}
\end{figure}

The other challenge with using the convolution layer is that only \emph{Conv2D} is currently supported by the Cerebras software stack, meaning that other convolution layers such as \emph{Conv3D} are not currently available for increased problem dimensions. Due to the ubiquity in HPC of PDEs in three dimensions, this omission would be a major limitation. To address this we increase the number of channels in the 2D convolution layer. Figure \ref{fig:convolution_layer_out_2d} illustrates the approach, where the number of channels in the convolution layer can be considered the depth of the stencil in the third dimension. Because the depth corresponds to the stencil size in the third dimension, as the filter slides across the input tensor in two dimensions each channel will undertake calculations on separate third dimension slices. However, as illustrated in Figure \ref{fig:convolution_layer_out_2d} this only results in a 2D output layer. To expand the number of output dimensions then the number of filter channels needs to be further increased by the number of input channels as illustrated by Figure \ref{fig:convolution_layer_out_3d}. This supports the handling of three dimensions, within the limitations imposed by the Cerebras software stack, but does imposes additional storage and computation overhead. 

\section{Results}
\label{sec:results}





In this section we conduct runs of our benchmark, a Jacobi method for solving Laplace's equation for diffusion in multiple dimensions, on the CS-1 which uses the latest version, 1.0.1, of Cerebras software. Performance is compared against \textbf{four} Nvidia Telsa V100-SXM2-16GB GPUs (CUDA toolkit version 10.1.243 and the CUDA library cuDNN version 7.6.5), and \textbf{two} 18-core Intel Xeon E5-2695 (Broadwell) CPUs. We use TensorFlow version 2.2.0 on the CS-1 and 2.3.0 on the GPUs and CPUs. Reported results are averaged over three runs.

To compare performance between the hardware we use the metric of \emph{delivered performance} in FLOPS. This is defined in Equation \ref{equation:delivered_flops}, where \emph{stencilFLOP} is the total number of floating point operations involved in applying the stencil for each output element. From the perspective of the computational algorithm this is the number of FLOPS delivered and includes the unnecessary floating point operations highlighted in Section \ref{sec:programming} which do not contribute to the final result. However there are additional internal operations being undertaken by the TensorFlow framework which are not readily discernible and these are not included in this metric. Consequently delivered performance can be thought of as a metric which is useful to compare the relative performance of hardware technologies, rather than able to provide an indication of absolute performance.

\begin{equation} delivered\;performance = (problemSize * stencilFLOP * iterations) / time
\label{equation:delivered_flops}
\end{equation}

As described in Section \ref{sec:programming}, the problem size is a product of $N$ and the number of steps, where $N$ is the size of the input tensor, for instance $X*Y$ in the 2D case. We set the batch size to be one, and the number of model iterations represents the number of solver iterations being undertaken, where an iteration operates on the data resulting from a previous iteration. 

\begin{table}[h]

 \centering
\begin{tabular}{ | c c c | }
\hline
\textbf{Technology} & \makecell{\textbf{Dense layer delivered} \\ \textbf{performance (GFLOPS)}} & \makecell{\textbf{Convolution layer delivered} \\ \textbf{performance (GFLOPS)}}\\\hline
Two CPUs (single precision) & 10.75 & 26.75\\
Two CPUs (mixed precision) & 0.63 & 3.88\\
Four GPUs (single precision) & 27.93 & 985.12\\
Four GPUs (mixed precision) & 32.28 & 1255.74\\
CS-1 (mixed precision) & 224.43 & 3054.89\\
\hline
\end{tabular}
\caption{Delivered performance for 2D Jacobi with a problem size of 2048 million elements ($X=Y=64$) using dense (over 7 iterations) and convolution (3500 iterations) layers across hardware and different numeric precision configurations}
\label{fig:performance}
\end{table}

Table \ref{fig:performance} reports the delivered performance in GFLOPS across the CPUs, GPUs, and Cerebras CS-1. On the CPUs and GPUs we include results for single and mixed precision (the later is a combination of 32-bit and 16-bit operations), whereas the Cerebras software stack only supports mixed precision for TensorFlow. For each of these configurations we include results for the dense and convolution layer approaches, with the dense layer running in \emph{training} mode and convolution layer in \emph{predict} mode. It is important to stress that the numbers reported here are delivered performance, for instance the GPU is capable of far higher raw FLOPS and the CS-1 was demonstrated to reach 0.86 PFLOPS in \cite{rocki2020fast}, however representing this benchmark in TensorFlow induces additional overhead and-so whilst this does not give a measure of the raw performance it does enable us to compare relative performance between the technologies.

It can be seen from the relative performance comparison in Table \ref{fig:performance} that the Cerebras CS-1 delivers around 2.5 times the performance of four V100 GPUs and around 114 times the performance of two 18-core Intel Xeon Platinum CPUs for this benchmark. \emph{Predict} mode, used for the convolution layer, is beneficial as the weights are already provided by the user and-so additional training work is not required. However not all TensorFlow operations support \emph{predict} mode on the WSE and the dense layer experiments can be run in \emph{train} mode only. 

Whilst our \emph{delivered performance} metric includes all stencil operations from the perspective of the algorithm, not all of these calculations are useful because not all contribute to the final result. For Laplace's equation for diffusion there are 7 useful calculations undertaken per input element, comprising four multiplications and three additions. However in the dense layer all input values contribute to each output element's calculation, resulting in $(N*2)-1$ operations for every output element. In the 2D case, with $X=Y=64$ and therefore $N=4096$, there are $8191$ operations for each output element and $33550336$ total calculations for the entire input tensor, per step, per iteration. The convolution layer by contrast undertakes 17 operations per output element, resulting in $69632$ total operations for the 2D case where $X=Y=64$. Whilst, as described in Section \ref{sec:programming}, there are $N*2$ additional operations for applying the mask with non-zero boundary conditions after an iteration, this is still considerably less overhead than the dense layer. The dense layer approach has a further limitation which is that a separate dense layer, of size $N^2$, must be created for each iteration. This limits the number of iterations with the dense layer to 7 on the CS-1, whereas the convolution approach can run at thousands of iterations. 



\begin{figure}[h]
\centering
\includegraphics[scale=0.55]{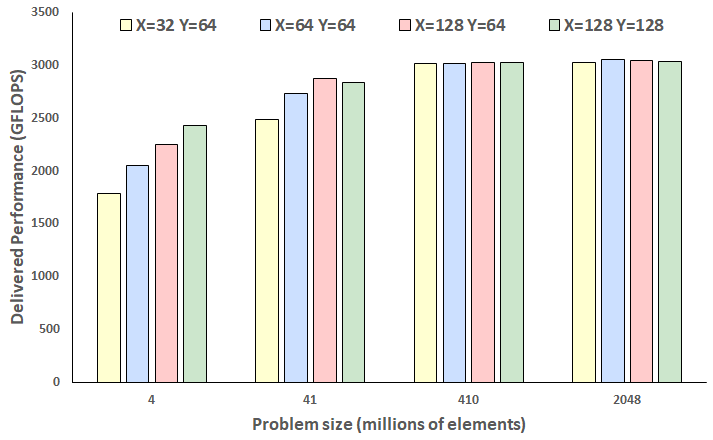}
\caption{Delivered performance for 2D Jacobi on the Cerebras CS-1 with 3500 iterations and 12 workers, with convolution layers. This experiment explores the performance impact for different problem sizes as the input tensor size and shape is varied}
\label{fig:nsize}
\end{figure}

Focusing on the convolution layer approach as it is more flexible and delivers much better performance convolution layer approach as it is more flexible and delivers much better performance we changed the size and shape of the input tensor from $X=Y=64$ that was used previously. Increasing the size and shape of the input tensor will result in a larger amount of input processed per step, consequently scaling the pipeline on the hardware to handle this and thus increasing the amount of fabric used on the WSE. Therefore it is interesting to see what difference this makes to performance, and Figure \ref{fig:nsize} illustrates the delivered performance in GFLOPS for four different problem size configurations where we modify the size and shape of the input tensor and the number of steps appropriately. It can be seen that this configuration change has an impact on performance at smaller problem sizes, where performance favours a larger input tensor processed per step and fewer steps. However as the problem size is increased the difference becomes smaller until, at 2048 million elements there is no significant difference between the configurations. The $32x64$ and $64x64$ shapes utilised 27\% of the CS-1 fabric, whereas the $128x64$ used 45\% and $128x128$ 67\%, beyond this size the Cerebras compiler was unable to find a suitable placement.

\begin{figure}[h]
\centering
\includegraphics[scale=0.45]{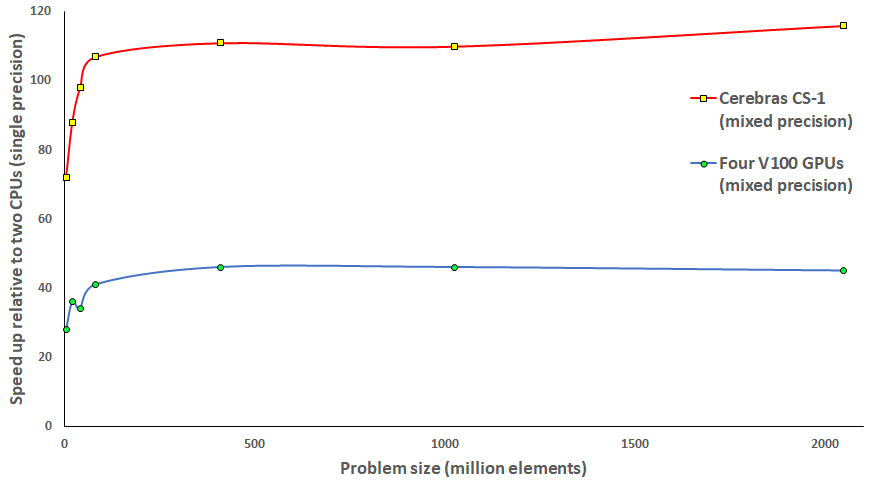}
\caption{Speed up relative to running single precision on two CPUs for 3D Jacobi. Using convolution layers, X=64, Y=64, Z=10, 3500 iterations, and 12 workers}
\label{fig:3djacobi}
\end{figure}

We then ran the 3D Jacobi benchmark with non-zero boundary conditions and an input tensor shape of $X=64, Y=64, Z=10$, which is the largest supported shape on the CS-1, with non-zero boundary conditions over 3500 iterations and 12 workers. Figure \ref{fig:3djacobi} reports the speed up obtained against a baseline of two 24-core Intel Xeon Platinum CPUs executing the benchmark in single precision (which as per Table \ref{fig:performance} is the best performing CPU configuration). We include results for four V100 GPUs at mixed precision, which is the highest performing GPU configuration, and the CS-1. It can be seen that the CS-1 significantly out-performs the CPUs and GPUs at all problem sizes, which broadly agrees with results reported for the 2D case in Table \ref{fig:performance}. It can be seen that speed up against the CPU is lower at smaller problem sizes for both the GPUs and Cerebras CS-1, although this is more pronounced for the CS-1, demonstrating that these accelerator technologies favour working on larger problem sizes and being fed with data to keep the fabric busy in the case of the CS-1.

\section{Conclusions}
\label{sec:conclusions}
The Cerebras Wafer Scale Engine (WSE) is an exciting technology which has already delivered significant advantages for machine learning. This makes it not only an important accelerator for AI, but also interesting for traditional computational HPC applications. In this paper we have explored the suitability of accelerating stencil-based computational algorithms on the WSE using TensorFlow via a benchmark which implements the Jacobi method for solving Laplace's equation for diffusion in multiple dimensions. This represents an important class of algorithm common place in HPC and-so insights gained are interesting for high performance workloads more widely. 

We ran performance experiments on a Cerebras CS-1, and because the exact operations being undertaken by the TensorFlow API are somewhat of a black-box, the \emph{delivered performance} metric was used which measures the performance delivered by the hardware from the perspective of the computational algorithm. This provides a relative, rather than absolute, measure of performance and enabled us to compare different hardware technologies. We found that, for this benchmark, the CS-1 delivered around two and a half times the performance of four V100 GPUs and 114 times the performance of two 18-core Intel Xeon Platinum (Broadwell) CPUs. 

Throughout this work we have found that the Cerebras CS-1 delivers very impressive performance, and whilst undoubtedly using TensorFlow to represent stencil-based computational algorithms is sub-optimal, this has provided us with the ability to undertake a relative performance comparison against other architectures and understand some of the behaviours of the WSE in more detail. The user experience in programming the WSE has been, in the main, pleasant and it is our belief that, given the performance results presented in this paper, it is very much worth the effort for HPC software developers to gain expertise with the Cerebras SDK \cite{sdk}.


%
%
%
\bibliographystyle{splncs04}
\bibliography{mybibliography}
%
\end{document}